**Stacking up electron-rich and electron-deficient monolayers to achieve extraordinary mid- to far-infrared excitonic absorption: Interlayer excitons in the C$_3$B/C$_3$N bilayer**

Zhao Tang,[1] Greis J. Cruz,[1] Fanhao Jia,[1,2] Yabei Wu,[3] Weiyi Xia,[4] and Peihong Zhang[1*]

[1]*Department of Physics, University at Buffalo, State University of New York, Buffalo, New York 14260, USA*

[2]*School of Materials Science and Engineering & International Centre of Quantum and Molecular Structures, Shanghai University, Shanghai 200444, China*

[3]*Department of Materials Science and Engineering, and Guangdong Provincial Key Lab for Computational Science and Materials Design, Southern University of Science and Technology, Shenzhen, Guangdong 518055, China*

[4]*Ames Laboratory, U.S. Department of Energy, Iowa State University, Ames, IA 50011, USA*

*E-mail: pzhang3@buffalo.edu

**Abstract**

Our ability to efficiently detect and generate far-infrared (i.e., terahertz) radiation is vital in areas spanning from biomedical imaging to interstellar spectroscopy. Despite decades of intense research, bridging the terahertz gap between electronics and optics remains a major challenge due to the lack of robust materials that can efficiently operate in this frequency range, and two-dimensional (2D) type-II heterostructures may be ideal candidates to fill this gap. Herein, using highly accurate many-body perturbation theory within the GW plus Bethe-Salpeter equation approach, we predict that a type-II heterostructure consisting of an electron rich $C_3N$ and an electron deficient $C_3B$ monolayers can give rise to extraordinary optical activities in the mid- to far-infrared range. $C_3N$ and $C_3B$ are two graphene-derived 2D materials that have attracted increasing research attention. Although both $C_3N$ and $C_3B$ monolayers are moderate gap 2D materials, and they only couple through the rather weak van der Waals interactions, the bilayer heterostructure surprisingly supports extremely bright, low-energy interlayer excitons with large binding energies of 0.2 ~ 0.4 eV, offering an ideal material with interlayer excitonic states for mid- to far-infrared applications at room temperature. We also investigate in detail the properties and formation mechanism of the inter- and intra-layer excitons.

## 1. Introduction

Far infrared (FIR) semiconductor detectors and emitters are of paramount importance in applications ranging from biomedical and thermal imaging [1], trace detection [2], to atmospheric and interstellar spectroscopy [3]. However, bridging the so-called terahertz (THz) gap has proven extremely challenging, in part due to the lack of suitable narrow gap semiconductors that can meet the stringent requirements of operations in the FIR/THz range. Atomically thin two-dimensional (2D) semiconductors, with their weak interlayer van der Waals (vdW) interactions, can be conveniently transferred and stacked together to form vertical heterostructures [4-6], adding another dimension to the materials design space that is otherwise inaccessible. 2D vdW heterostructures provide practically unlimited possibilities of creating new materials and structures, offering a unique platform for studying new excitonic physics [7].

Of particular interest are 2D heterostructures with a type-II band alignment [8-20]. Not only may these heterostructures enable ultrafast electron-hole (*e-h*) separation and charge transfer after optical excitations [14], but they may also facilitate the formation of spatially-indirect interlayer excitons with excitation energies that are far smaller than those of the individual layers, paving the way of designing infrared materials using 2D semiconductors with moderate band gaps [16]. So far, much research has been focusing on type-II heterostructures made from transition metal dichalcogenides (TMDs) [5, 8-16, 21]. However, optical absorption arising from interlayer excitons are often relatively weak [22] (compared with intralayer excitons) in TMD-based heterostructures, presumably due to the minimal overlap between the electron and hole states which are mostly derived from the *d* states of the transition metals in opposite layers, thus limiting their applications as absorbers using interlayer excitons.

In this work, using many-body perturbation theory (MBPT) calculations within the GW plus Bethe-Salpeter equation (BSE) approach, we show that the bilayer heterostructure constructed using ordered carbon-nitrogen and carbon-boron 2D alloys $C_3N$ [23-26] and $C_3B$ [26-30], two remarkably stable 2D semiconductors with moderate band gaps, gives rise to surprisingly strong interlayer excitonic absorption in the mid-infrared (MIR) to FIR range, peaking at about 0.18 eV and extending to as low as 40 meV. Whereas $C_3B$ is electron deficient (compared with graphene), $C_3N$ is electron rich. Thus, they form an ideal pair for constructing type-II heterostructures [31]. The primarily $p_z$ orbitals derived band edge states allow significant overlap between the electron and hole states of the bilayer system, resulting in very large dipole transition matrix elements for the interlayer excitons. Detailed analyses of the state-dependent exciton binding energy reveal a shell-like distribution of the excitonic states. In addition to the conventional assignments of inter- and intra-layer excitons, we observe excitonic states that have strongly hybridized inter- and intra-layer components. The exciton binding energies of the bilayer system, although generally smaller than those of monolayer systems, are still significant (0.2 ~ 0.4 eV) and far greater than those of conventional narrow gap semiconductors (typically of the order of meV), providing robust interlayer excitonic states for MIR to FIR applications at room temperature.

## 2. Computational details

Structural optimizations are carried out using the van der Waals (vdW) functional optB86b [32] within the density functional theory (DFT) as implemented in the Quantum Espresso package [33, 34]. A local version of the BerkeleyGW code [35, 36] is used for carrying out the GW [37] and BSE [36] calculations. The recently developed acceleration methods [38, 39] are used for the GW calculations, which lead to a combined speed-up factor of over 1,000 for 2D materials. We have carefully checked the convergence of our GW and BSE calculations. We include a large vacuum layer of 40 a.u. and use a slab-truncated Coulomb potential [40] in our calculations to reduce the interaction from the periodic images. The Hybertsen-Louie Generalized Plasmon-Pole (HL-GPP) model [37] is used to extend the static dielectric function to finite frequencies. A cutoff energy of 60 Ry is used for the DFT pseudopotential plane-wave calculations, and a high kinetic cutoff of 40 Ry is used for the dielectric matrices. A dual-grid method [36] is applied to reduce the workload of the BSE calculations: The $e$-$h$ kernel is first calculated on an 18×18×1 coarse $k$-grid, the results are then interpolated onto a 60×60×1 fine $k$-grid. Since this system has a relatively large unit cell, the fine $k$-grid used here is equivalent to a 120×120×1 grid for a two-atom graphene or hexagonal BN unit cell.

## 3. Results and discussion

### A. Crystal structure and quasiparticle properties of the $C_3B$/$C_3N$ bilayer

The $C_3N$ ($C_3B$) monolayer can be viewed as graphene with 25% of its atoms replaced by nitrogen (boron), as shown in Fig. 1(a). In the lowest-energy $C_3N$/$C_3B$ bilayer structure [Fig. 1(a)], half of the N atoms are on top of B; the other half are on top of the center of the hexagons. This optimized structure is consistent with previous studies [31, 41]. The optimized interlayer separation within the optB86b-vdW functional [32] is about 3.16 Å, with an interlayer binding energy of about 44 meV/atom. Since $C_3N$ is electron rich (compared with graphene) and $C_3B$ is electron deficient, they may form an ideal pair for constructing type-II 2D heterostructures. The $C_3N$/$C_3B$ bilayer [Fig. 1(a)] may also help overcome the issue of weak interlayer excitons absorption [22] observed in TMD-based heterostructures since the valence and conduction edges of both $C_3N$ and $C_3B$ are primarily derived from the out-of-plane $p_z$ orbitals, which can lead to significant overlap between the electron and hole states, thus potentially strong interlayer optical absorption.

The quasiparticle and optical properties of the monolayer systems have been discussed in detail elsewhere [25, 26, 30], here we summarize a few main features. $C_3N$ and $C_3B$ are both indirect-gap semiconductors as shown in Fig. 1(b) and (c) in which the DFT and GW band structures are shown with orange dashed and black solid curves, respectively. Note that the quasiparticle band structure [Fig. 1 (d)] of the $C_3N$/$C_3B$ bilayer is not just a simple superposition of those of the individual monolayers as a result of combined chemical hybridization and interlayer screening effects. For example, the calculated quasiparticle excitation gap at the M point of the bilayer system with predominantly $C_3N$ character is 2.05 eV, to be compared with 2.74 eV for the monolayer $C_3N$. We also list in Table I other noticeable changes in the calculated quasiparticle excitation gaps.

Both monolayer systems show extremely strong and narrow excitonic absorption peaks in the visible region (1.9 eV for $C_3N$ and 2.1 eV for $C_3B$) [26] due to the presence of nearly parallel valence and conduction bands [Fig. 1(b) and (c)]. Whereas the valence band maximum (VBM) of the $C_3N$ monolayer locates at the M point, for $C_3B$, it is the conduction band minimum (CBM) that locates at the M point. As a result, when $C_3N$ and $C_3B$ are stacked together, the band structure displays a nearly-perfect *e-h* symmetry, as shown in Fig. 1 (d). Perhaps more important is the formation of a very small gap near the M point of the $C_3N/C_3B$ bilayer. The calculated direct gap at the M point is 0.17 eV at the DFT level. This value increases to 0.33 eV when the quasiparticle corrections are included within the GW approximation. If the optical transition between the CBM and VBM around the M point is allowed, it would make the $C_3N/C_3B$ bilayer a very promising type-II heterostructure for infrared applications. A simple analysis of the atomic characters of the band edge states would help shed some light on the optical properties of this material.

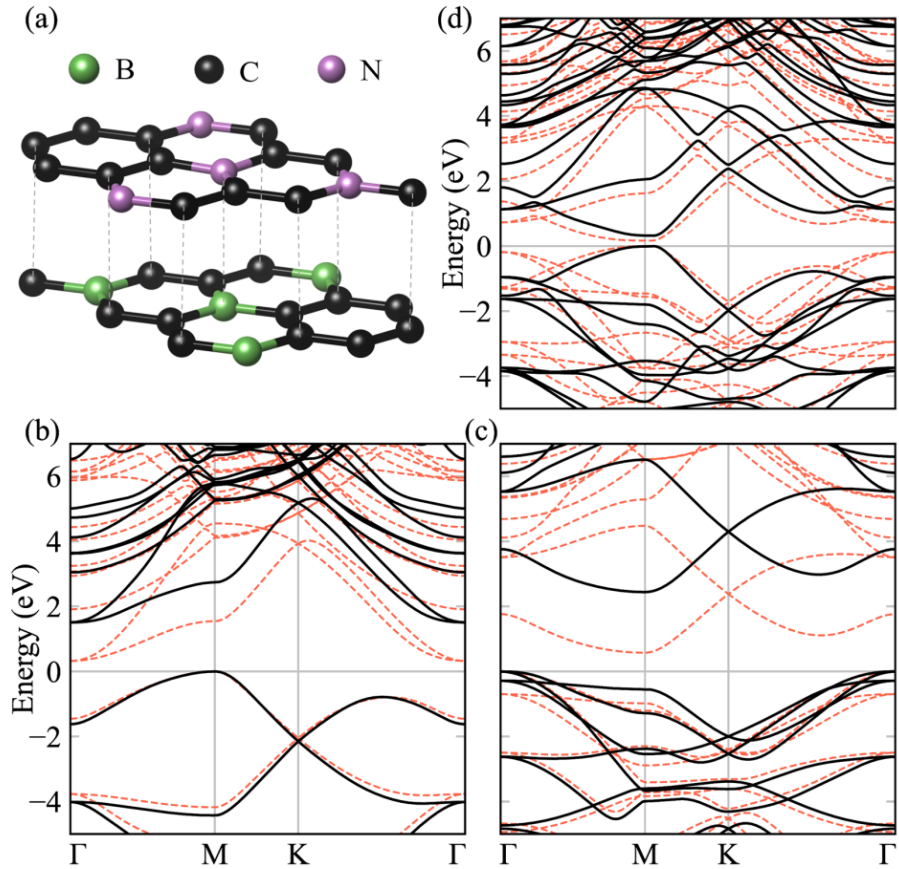

FIG. 1. Crystal structure and quasiparticle band structures. (a) The optimized structure of the $C_3N/C_3B$ bilayer. (b ~ d) The band structures of $C_3N$ monolayer (b), $C_3B$ monolayer (c), and $C_3N/C_3B$ bilayer (d). The DFT results are shown with dashed orange, and the GW results are show with solid black.

TABLE I. Quasiparticle excitation gaps (in eV) at the Γ and M points of the monolayer and bilayer systems.

| Gap | C$_3$N-mono | C$_3$B-mono | C$_3$N@bilayer | C$_3$B@bilayer |
|---|---|---|---|---|
| Γ | 3.13 | 3.76 | 2.65 | 2.76 |
| M | 2.74 | 2.98 | 2.05 | 2.11 |

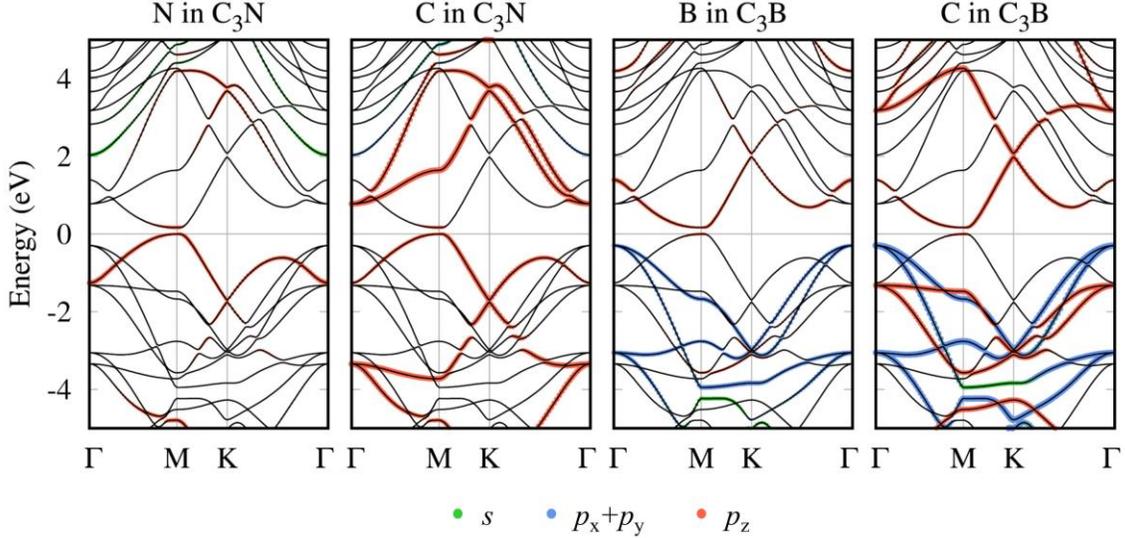

FIG. 2. Decomposition of the band wave functions into contributions from atomic orbitals. The contributions from $s$, $p_x+p_y$, and $p_z$ orbitals are colored in green, blue, and orange, respectively.

Figure 2 shows the decomposition of the band (Bloch) wave function onto contributions from atomic orbitals. The bilayer system indeed forms a type-II band alignment, where the VBM and CBM are derived mainly from the C$_3$N and C$_3$B layers, respectively. However, interlayer hybridization can also be clearly seen. When *e-h* excitations occur across the band gap, the electron would reside in the C$_3$B layer, whereas hole would be in the C$_3$N layer, forming spatially separated interlayer excitons. Another important feature is that the band edge states are primarily of $p_z$ character. These orbitals protrude into the inter-layer region, thus are potentially beneficial for strong optical absorption.

## B. Excitonic structure and optical properties

We now investigate the *e-h* excitations and optical properties of the C$_3$N/C$_3$B bilayer by solving the BSE equation within the Tamm-Dancoff approximation [36, 42]:

$$\left(E_{c\mathbf{k}} - E_{v\mathbf{k}}\right) A^S_{vc\mathbf{k}} + \sum_{v'c'\mathbf{k}'} \langle vc\mathbf{k}|K^{\text{eh}}|v'c'\mathbf{k}'\rangle A^S_{v'c'\mathbf{k}'} = \Omega^S A^S_{vc\mathbf{k}}, \qquad 1$$

where $E_{c\mathbf{k}}$ and $E_{v\mathbf{k}}$ are the quasiparticle energies of the conduction and valence states calculated within the GW approximation, and $K^{\text{eh}}$ is the *e-h* interaction kernel. Solving the above eigenvalue problem gives the *e-h* excitation energies $\Omega^S$ and wave functions:

$$\Psi^S(\boldsymbol{r}_e, \boldsymbol{r}_h) = \sum_{vc\boldsymbol{k}} A^S_{vc\boldsymbol{k}} \psi_{c\boldsymbol{k}}(\boldsymbol{r}_e) \psi^*_{v\boldsymbol{k}}(\boldsymbol{r}_h). \qquad (2)$$

The imaginary part of the macroscopic transverse dielectric function is then given by

$$\epsilon_2(\omega) = \frac{16\pi^2 e^2}{\omega^2} \sum_S |\mathrm{e} \cdot \langle 0|v|S\rangle|^2 \delta(\omega - \Omega^S), \qquad (3)$$

where $v$ is the velocity operator, e is the polarization vector of the light, and $\omega$ is the energy of the photon.

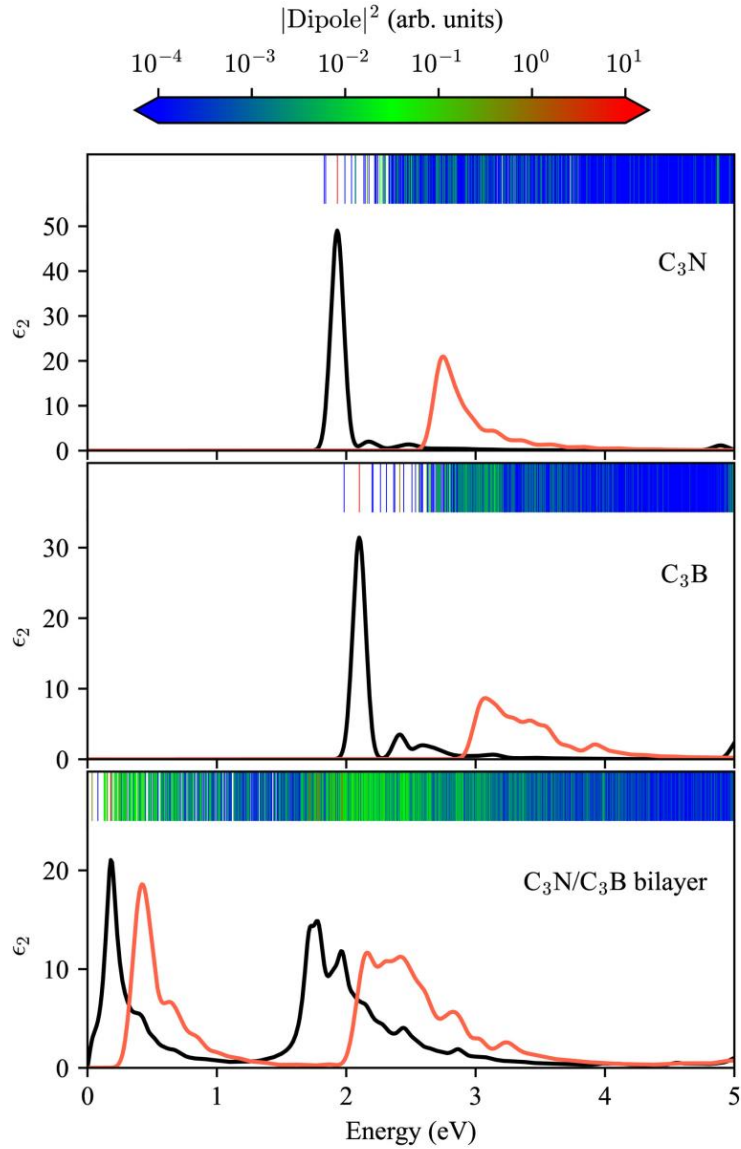

FIG. 3. The imaginary part of the dielectric function of the $C_3N$ monolayer (top), $C_3B$ monolayer (middle), and $C_3N/C_3B$ bilayer (bottom). The BSE results are shown in black; the GW results are shown in orange. A Gaussian broadening of 0.05 eV is used in the calculations.

Figure 3 compares the optical absorption (i.e., the imaginary part of the dielectric function $\epsilon_2(\omega)$) of $C_3N$ (top panel) and $C_3B$ (middle) monolayers and $C_3N/C_3B$ bilayer (bottom) below 5 eV. The results calculated without the *e-h* interaction (i.e., at the GW level) are shown in orange, and those with the *e-h* interaction are shown in black. The exciton states of each system, color coded by their optical transition matrix elements, are shown as vertical lines on top of each $\epsilon_2(\omega)$ plot. Both $C_3N$ and $C_3B$ monolayers show extremely strong and narrow absorption [26] in the visible range arising from the transitions between the nearly parallel conduction and valence bands as mentioned earlier. In the bilayer system, the intralayer excitonic absorption peaks can still be recognized, although they are smeared and shifted to lower energies due to the interlayer coupling and screening effects. The strength of these intralayer exciton absorption peaks in the bilayer system are also considerably weaker than those in the individual monolayer systems. The most striking feature, however, is the emergence of very strong absorption in the infrared region, which is absent in the monolayers and clearly is a result of interlayer excitonic absorption. This interlayer absorption peaks at around 0.18 eV, and it is even stronger than the intralayer absorption, thus providing an ideal material for the sought-after mid- to far-infrared applications [20].

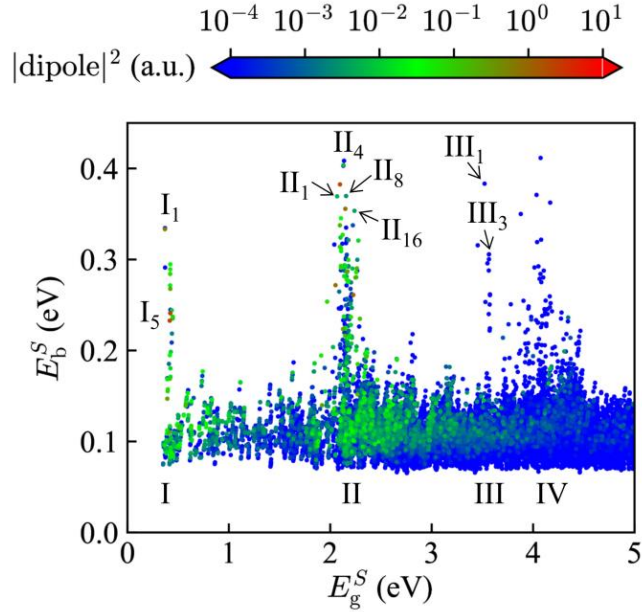

FIG 4. Excitonic structure of $C_3N/C_3B$ bilayer showing by plotting the exciton binding energy $E_b^S$ vs. the non-interacting excitation energy $E_g^S$. Excitonic states with large exciton binding energies exhibit a shell-like distribution. The data are colored according to the optical dipole transition matrix elements of the states. A few selected states are labeled for later discussion.

In order to gain better insight into the excitonic structure of the $C_3N/C_3B$ bilayer, we investigate the exciton binding energies, which is defined as the difference between the interacting and non-interacting *e-h* excitation energies [26]:

$$E_b^S = E_g^S - \Omega^S, \qquad 4$$

where the non-interacting excitation energy $E_g^S$ of a given excitonic state $|S\rangle$ is defined using the *e-h* amplitude $A_{vc\bm{k}}^S$ as

$$E_g^S = \sum_{vc\bm{k}} |A_{vc\bm{k}}^S|^2 \; E_{c\bm{k}} - E_{v\bm{k}} \; . \qquad 5$$

Using this definition, we can calculate the binding energy of any exciton states. Figure 4 shows the exciton binding energy $E_b^S$ vs. $E_g^S$ for the bilayer system. Most states have rather small binding energies between 0.07 and 0.15 eV. The states with large binding energies ($> 0.2$ eV) group into series and display an interesting shell-like distribution. Series I excitons are clearly interlayer excitons, whereas series II are a mixture of inter- and intra-layer excitons, as we will discuss in more details later. The binding energies of these interlayer excitons are in general slightly smaller than those of the II series. Compared with the monolayer $C_3N$ and $C_3B$ [26], the exciton binding energies in the bilayer system are significantly reduced. The largest binding energy is only about 0.4 eV for the bilayer system, to be compared with over 1 eV for the monolayers. This significantly reduced binding energy is a combined effect of stronger dielectric screening and more extended *e-h* wave functions in the bilayer system. Table II lists important properties of a few low energy interlayer excitons (the I series) including the degree of degeneracy and the optical dipole transition moment. Except for the $I_2$ exciton which has a negligible optical transition matrix element, all other low energy interlayer excitons are bright; the interlayer optical absorption peaks around 0.18 eV and extends to below 40 meV.

TABLE II. Low energy I-series excitonic states. Shown in the table are the degree of degeneracy (Deg.), exciton energy ($\Omega^S$), quasiparticle excitation energy ($E_g^S$), exciton binding energy ($E_b^S$) and the optical dipole matrix elements. In cases that the excitonic states are nearly degenerate, they are denoted as ND. All energies are in electron-Volts (eV).

| Interlayer exciton | Deg. | $\Omega^S$ | $E_g^S$ | $E_b^S$ | $|\text{dipole}|^2$ (a.u.) |
|---|---|---|---|---|---|
| $I_1$ | 2 | 0.036 | 0.370 | 0.334 | $6.34\times10^{-1}$ |
| $I_2$ | 1 | 0.080 | 0.371 | 0.291 | $5.94\times10^{-5}$ |
| $I_3$ | 3 (ND) | 0.132 | 0.422 | 0.289 | $1.55\times10^{-1}$ |
| $I_4$ | 2 | 0.152 | 0.421 | 0.270 | $2.58\times10^{-1}$ |
| $I_5$ | 3 (ND) | 0.181 | 0.421 | 0.239 | $4.96\times10^{0}$ |
| $I_6$ | 1 | 0.192 | 0.437 | 0.244 | $1.12\times10^{-2}$ |
| $I_7$ | 2 | 0.210 | 0.446 | 0.236 | $9.97\times10^{-2}$ |

## C. Formation of the inter- and intra-layer excitons

The formation and characteristics of the excitonic states can be better illustrated using the *e-h* amplitude functions $A_{vc\bm{k}}^S$ defined in Eq. (1) and (2). Visualizing this multidimensional function, however, is often challenging. To this end, we define a *k*-dependent electron amplitude function,

$$|A^S_{c\bm{k}}|^2 = \sum_v |A^S_{vc\bm{k}}|^2, \qquad 6$$

which essentially shows the contributions from the electron (conduction) states to a given exciton $S$, the hole amplitude,

$$|A^S_{v\bm{k}}|^2 = \sum_c |A^S_{vc\bm{k}}|^2, \qquad 7$$

and the pair amplitude,

$$|A^S_{\bm{k}}|^2 = \sum_{vc} |A^S_{vc\bm{k}}|^2. \qquad 8$$

These functions help reveal how the (non-interacting) electron and hole states superimpose to form inter- and intra-layer excitons.

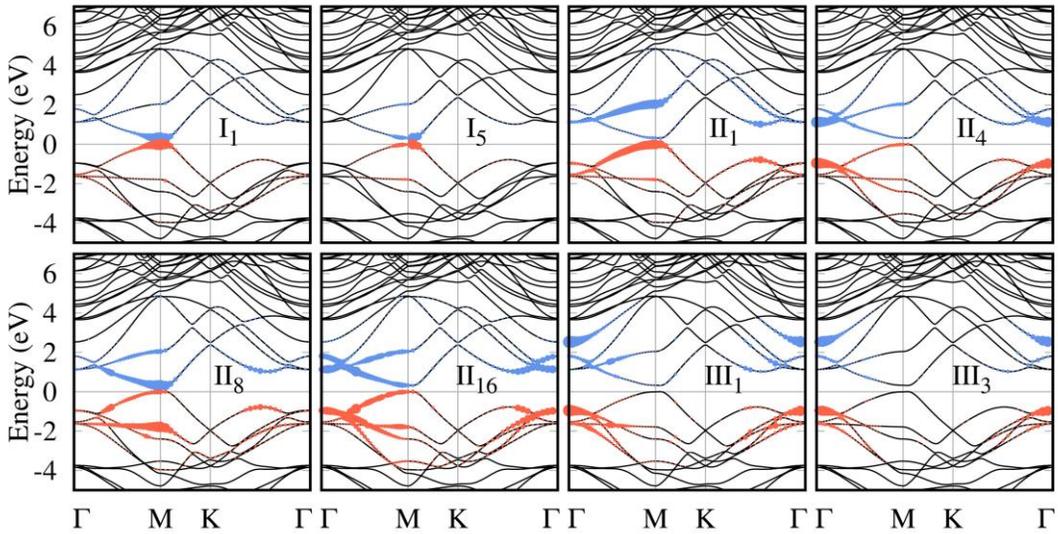

FIG. 5. Electron (shown in blue) and hole (orange) amplitudes of a few selected low-energy excitons as marked in Fig. 4.

Figure 5 shows the *k*-dependent *e/h* amplitudes of several selected low-energy excitons from series I, II, and III, as marked in Fig. 4. The I-series excitons are clearly interlayer excitons, mainly comprising electron states near the CBM from the $C_3B$ layer and hole states near the VBM from the $C_3N$ layer. Interestingly, the $I_5$ exciton is nearly 10 times brighter than the $I_1$ exciton as shown in Table II. The excitation and binding energies of the I-series excitons can be found in Table II. The $II_1$ exciton ($\Omega^S = 2.07$ eV; $E^S_b = 0.37$ eV) can be identified as an intralayer exciton in the $C_3N$ layer. The $II_4$ exciton ($\Omega^S = 2.13$ eV; $E^S_b = 0.41$ eV), on the other hand, is an interlayer exciton since the electrons mostly reside in the $C_3N$ layer, whereas holes in the $C_3B$ layer. The $II_8$ and $II_{16}$ states can be characterized as intralayer excitons, but with intralayer components from both the $C_3N$ and $C_3B$ layers. Both $III_1$ and $III_3$ are interlayer excitons with hole primarily localized in the $C_3N$ layer and electron in the $C_3B$ layer. Strictly speaking, however, there are no pure intra-or inter-layer excitons in a bilayer (or multilayer) system. In other words, all excitonic states always have both inter- and intra-layer components. Finally, we show in Fig. 6 the $k$-dependent pair amplitude defined in Eq. 8 corresponding to the states shown in Fig. 5. These plots better illustrate the distribution of the $e$-$h$ amplitudes in the entire Brillouin zone. For example, states $I_1$, $I_5$, and $II_8$ are highly localized at or near the M point. State $II_1$ mainly comprises transitions along the M−Γ direction, whereas states $II_4$, $II_{16}$, $III_1$, and $III_3$ are heavily localized near the Γ point.

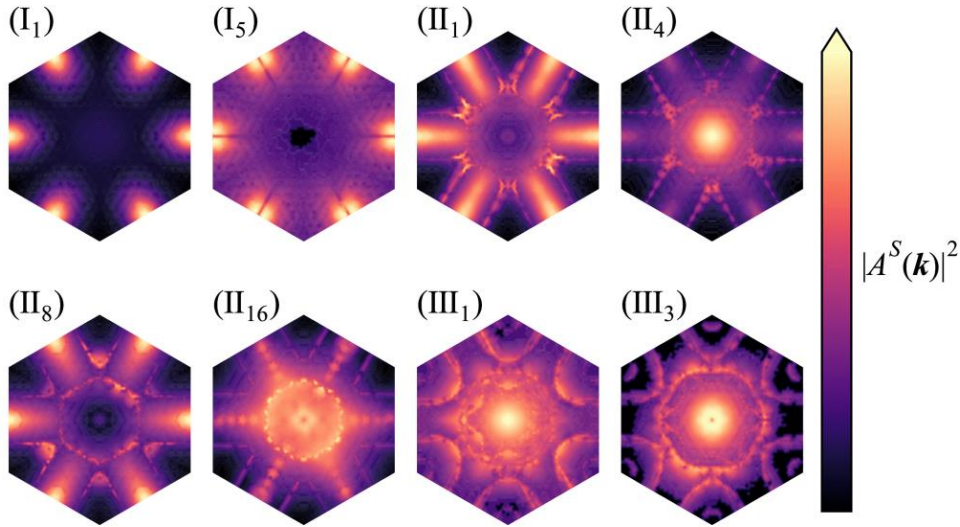

FIG. 6. Contour plots of $k$-dependent $e$-$h$ pair amplitude in the Brillouin zone of selected low-energy excitons from I, II, and III series as marked in Fig. 4. For degenerate states, we average their amplitudes to show full symmetry.

## 4. Summary

The weak vdW interaction between 2D materials enables near-effortless transfer and construction of vertically stacked dissimilar layers to form heterostructures. These structures may combine electronically and/or structurally disparate layers into single functional materials, offering a unique dimension to the materials design space that is otherwise inaccessible. Perhaps more important (and somewhat surprising) is that, although the interlayer coupling may be rather weak in the electronic ground state, the interaction between the excited electrons and holes residing in different layers can be strong and optically active. Using highly accurate MBPT calculations, we have shown that stacking up electron rich $C_3N$ and electron deficient $C_3B$ monolayers results in a type-II heterostructure and the formation of interlayer excitons with exceptionally strong optical activities in the mid- to far-infrared range. The large binding energy (0.2 ~ 0.4 eV) of the interlayer excitons ensures stable operations of exciton-based optoelectronics in the mid- to far-infrared range at or above room temperature.

**Data availability statement**

The data that support the findings of this study are available upon reasonable request from the authors.

**Acknowledgements**

This work is supported by the National Science Foundation under Grants No. DMREF-1626967. Work at SUSTECH and SHU is supported by the National Natural Science Foundation of China (Grant No. 12104207 and 11929401). We acknowledge computational support from the Center for Computational Research, University at Buffalo.

**Conflict of interest**

The authors declare no competing interests.